\documentclass[preprint,12pt]{elsarticle}

\usepackage{epsfig}

\usepackage{amssymb}
\usepackage{amsmath}

\begin{document}

\begin{frontmatter}

%% Title, authors and addresses

%% use the tnoteref command within \title for footnotes;
%% use the tnotetext command for the associated footnote;
%% use the fnref command within \author or \address for footnotes;
%% use the fntext command for the associated footnote;
%% use the corref command within \author for corresponding author footnotes;
%% use the cortext command for the associated footnote;
%% use the ead command for the email address,
%% and the form \ead[url] for the home page:
%%
%% \title{Title\tnoteref{label1}}
%% \tnotetext[label1]{}
%% \author{Name\corref{cor1}\fnref{label2}}
%% \ead{email address}
%% \ead[url]{home page}
%% \fntext[label2]{}
%% \cortext[cor1]{}
%% \address{Address\fnref{label3}}
%% \fntext[label3]{}

\title{Universal Quantum Circuit of Near-Trivial Transformations}

%% use optional labels to link authors explicitly to addresses:
%% \author[label1,label2]{<author name>}
%% \address[label1]{<address>}
%% \address[label2]{<address>}

\author{Min Liang}
\author{Li Yang}
\ead{yangli@gucas.ac.cn}
\address{State Key Laboratory of Information
Security, Graduate University of Chinese Academy of Sciences, Beijing 100049, China}

\begin{abstract}
%% Text of abstract
Any unitary transformation can be decomposed into a product of a group of near-trivial transformations. We investigate in details the construction of universal quantum circuit of near trivial transformations. We first construct two universal quantum circuits which can implement any single-qubit rotation $R_y(\theta)$ and $R_z(\theta)$ within any given precision, and then we construct universal quantum circuit implementing any single-qubit transformation within any given precision. Finally, a universal quantum circuit implementing any $n$-qubit near-trivial transformation is constructed using the universal quantum circuits of $R_y(\theta)$ and $R_z(\theta)$. In the universal quantum circuit presented, each quantum transformation is encoded to a bit string which is used as ancillary inputs. The output of the circuit consists of the related bit string and the result of near-trivial transformation. Our result may be useful for the design of universal quantum computer in the future.
\end{abstract}

\begin{keyword}
%% keywords here, in the form: keyword \sep keyword

%% MSC codes here, in the form: \MSC code \sep code
%% or \MSC[2008] code \sep code (2000 is the default)
quantum computation \sep quantum circuit \sep universal quantum circuit \sep near-trivial transformation
\end{keyword}

\end{frontmatter}

%%
%% Start line numbering here if you want
%%
% \linenumbers

%% main text

\section{Introduction}
Deutsch proved that any $d$-dimensional unitary transformation can be
decomposed into a product of $2d^2-d$ two-level unitary transformations~\cite{Deutsch89,Ekert96}.
Bernstein and Vazirani~\cite{Bernstein97} proved that any unitary transformation can be decomposed into a product of near-trivial transformations which is a special case of two-level unitary transformation. Then, based on near-trivial transformation, we can perform any quantum operation on quantum states.

DiVincenzo~\cite{DiVincenzo95} proposed a constructive way in which
any two-level unitary transformation could be implemented using CNOT and single-qubit gates.
Since near-trivial transformation is a kind of two-level unitary transformation,
any near-trivial transformation could be implemented in the way of DiVincenzo.
However, it is not a universal implementation because we have to know which dimensions
the transformation performs on before implementing the transformation.

There are various researches about the construction of quantum circuit implementing general unitary transformation, such as Khaneja-Glaser decomposition(KGD)\cite{Khaneja01}, Cosine-sine decomposition(CSD)\cite{Paige94}, QR decomposition \cite{Barenco95}, and so on. By means of KGD, some constructive ways for general $2$-qubit and $3$-qubit gates were
proposed~\cite{Zhang03,Vidal04,Vatan04,Vatan04-2,Wei08}. According to~\cite{Mottonen04}, a quantum circuit for general $n$-qubit gate can
be constructed uniformly based on CSD. Based on Cartan's KAK decomposition(CSD and KGD are special
cases of KAK decomposition), a new way was shown to uniformly produce quantum
circuit implementing arbitrary unitary transformation \cite{Nakajima05}.

Bullock and Markov~\cite{Bullock03} proposed a recursive and constructive way to uniformly
construct asymptotically optimal circuit for arbitrary $n$-qubit diagonal unitary transformation.
Liu et al.\cite{Liu08} suggested a method to uniformly construct a polynomial-size quantum circuit for general
$n$-qubit controlled unitary transformation. Sousa and
Ramos~\cite{Sousa07} proposed a universal cell consists of single-qubit gates with adjustable parameters and CNOT gates that can be switched to an identity gate. By setting these parameters, it can perform CNOT on any two qubits, or arbitrary single-qubit gate on any qubit of the $n$ qubits. Because any unitary gate can be decomposed into CNOT and single-qubit gates, the products of several universal cells can implement any unitary transformation. Besides, Long et al.~\cite{Long09} firstly introduced allowable generalized quantum gates and its realization was shown by Zhang et al.~\cite{Zhang10}.

In this paper, we construct universal quantum circuit which is in
the standard form of quantum circuit. Firstly, based on a set of
discrete gates containing CNOT and single-qubit gates, we
construct two universal quantum circuits which can implement any
single-qubit rotation $R_y(\theta)$ and $R_z(\theta)$ within any given
precision. Furthermore, we also construct universal quantum circuit
implementing any single-qubit transformation within any given
precision. Bernstein and Vazirani~\cite{Bernstein97} proposed the
idea for constructing a quantum Turing machine(QTM) which can carry
out near-trivial rotation. Based on this, we construct a
quantum circuit which can exactly implement near-trivial
transformation. Finally, a universal quantum circuit which can
implement any $n$-qubit near-trivial transformation is constructed
using the universal quantum circuits of $R_y(\theta)$ and
$R_z(\theta)$. Compared with the programmable quantum circuit in
Sousa and Ramos~\cite{Sousa07}, our construction of universal
quantum circuit is in the standard form of quantum circuit. In our
universal quantum circuit, each quantum transformation is relative
to a bit string. The bit string is input as ancillary bits, then the
circuit will carry out that quantum transformation. Our construction consists of only definitive quantum gates and can be easily simulated by QTM. Compared with other constructions of universal quantum circuit, our construction is the only one which can be changed directly to get an universal QTM.

Since all the gates in our construction are fixed gates, it opens the possibility to implement
exponential number of near-trivial transformations in a single circuit. Since any unitary transformation
can be decomposed into a product of some near-trivial transformations, our result may contribute to the design of universal quantum computer.
\section{Preliminaries}
\subsection{Some notations and definitions}
\noindent
Let $C^n_{b_1b_2\cdots b_n}(U)$  denote a $n+1$-qubit gate with a transformation $U$ being performed on the last qubit,
conditional on the $n$ control qubits being set to $b_1b_2\cdots b_n$, where $U$ is a $1$-qubit unitary operation.
When $U=\left(\begin{array}{cc}0 & 1 \\ 1 & 0 \end{array}\right)$, then $C^n_{b_1b_2\cdots b_n}(U)$ is a $n+1$-qubit
generalized Toffoli gate. For convenience, $C^n_{11\cdots 1}(U)$ is denoted as $C^n(U)$. If $n=2$ and $U=\left(\begin{array}{cc}0 & 1 \\ 1 & 0 \end{array}\right)$, $C^n(U)$ is the Toffoli gate.

$M$ is a $d$-dimensional complex matrix. Let $M_{jk}$ denote the entry in row $j$ and column $k$. Let $e_i$ denote
a unit-length column vector with only the  $i$th entry being $1$.

Near-trivial transformation is a class of exceedingly simple unitary transformations. These transformations
either apply a phase shift in one dimension or apply a rotation between two dimensions, while act as
the identity otherwise. The definition is as follows:

{\bf Definition~1(Bernstein and Vazirani~\cite{Bernstein97}):} A unitary matrix M is near-trivial if it satisfies one of the following two conditions.
%\begin{romanlist}
% \item $M$ is the identity except that one of its diagonal entries is $e^{i\theta}$ for some $\theta\in[0,2\pi]$.For example,$\exists j$,$M_{jj}=e^{i\theta} \forall k\neq j M_{kk}=1$,and $\forall k\neq l M_{kl}=0$.
% \item $M$is the identity except that the submatrix in one pair of distinct dimensions $j$ and $k$ is the rotation by some angle $\theta\in[0,2\pi]:(\begin{array}{cc}cos\theta & -sin\theta \\ sin\theta & cos\theta )$. So, as a transformation $M$ is near-trivial if there exists $\theta$ and $i\neq j$ such that $Me_i=(cos\theta)e_i+(sin\theta)e_j$,$Me_j=(-sin\theta)e_i+(cos\theta)e_j$,and $\forall k\neq i,j Me_k=e_k$.
%\end{romanlist}
\begin{enumerate}
 \item $M$ is the identity except that one of its diagonal entries is $e^{i\theta}$ for some $\theta\in[0,2\pi]$. For example, $\exists j,M_{jj}=e^{i\theta}$,$\forall k\neq j,M_{kk}=1$, and $\forall k\neq l,M_{kl}=0$.
 \item $M$ is the identity except that the submatrix in one pair of distinct dimensions $j$ and $k$ is the rotation by some angle $\theta\in[0,2\pi]$:$\left(\begin{array}{cc}cos\theta & -sin\theta \\ sin\theta & cos\theta\end{array}\right)$. So, as a transformation $M$ is near-trivial if there exists $\theta$ and $i\neq j$ such that $Me_i=(cos\theta)e_i+(sin\theta)e_j$,$Me_j=(-sin\theta)e_i+(cos\theta)e_j$, and $\forall k\neq i,j,~Me_k=e_k$.
\end{enumerate}

We call a transformation which satisfies statement $1$ a near-trivial phase shift denoted as $[j,j,\theta]$, and we call a transformation which satisfies statement $2$ a near-trivial rotation denoted as $[i,j,\theta]$,where $i\neq j$.

We use a unified notation $[x,y,\theta,\theta']$ for near-trivial transformations. The unified notation $[x,y,\theta,\theta']$ represents a near-trivial rotation by an angle $\theta$ between dimensions $x$ and $y$  while $x\neq y$ , and a near-trivial phase shift of $e^{i\theta'}$  in dimension $x$  while $x=y$. In other words, $[x,y,\theta,\theta']=[x,y,\theta]$, if $x\neq y$ and $[x,y,\theta,\theta']=[x,x,\theta']$, if $x=y$.

Let $R(\theta)$ denote the matrix $\left(\begin{array}{cc}cos\theta & -sin\theta \\ sin\theta & cos\theta\end{array}\right)$ and $P(\theta')$  denote the matrix $\left(\begin{array}{cc}1&0\\0& e^{i\theta'}\end{array}\right)$.

Let $R_x(\theta),R_y(\theta),R_z(\theta)$ denote the rotation by $\theta$ around the axis $\hat{x}$,$\hat{y}$,$\hat{z}$ respectively.

{\bf Definition~2(Bera et al.~\cite{Bera08}):} Fix $n>0$ and let $\mathcal{U}$ be a collection of unitary transformations on $n$ qubits. A quantum circuit $C_u$ on $n+m$ qubits is universal for $\mathcal{U}$ if, for each transformation $U\in\mathcal{U}$, there is a string $e\in\{0,1\}^m$(the encoding) such that for all strings $d\in\{0,1\}^n$(the data),
\begin{equation}C_u(|e\rangle \otimes |d\rangle) =  |e\rangle \otimes(U|d\rangle ).\end{equation}

\subsection{DiVincenzo's implementation of near-trivial transformation}

In the following part, both $x$ and $y$ are arbitrary $n$-bit strings.

A Gray code connecting $x$ and $y$ is a sequence of bit strings, starting with $x$ and concluding with $y$, such that adjacent bit strings differ only by a single bit. According to the Gray code connecting $x$ and $y$, we can construct a quantum circuit implementing a two-level unitary transformation between dimensions $x$ and $y$~\cite{Nielsen02}.

Generally, the number of Gray codes connecting $x$ and $y$ is not less than $1$. Denote the Hamming distance of $x$ and $y$ as $Ham(x,y)$, each Gray code connecting $x$ and $y$ has $Ham(x,y)+1$ elements.

We briefly describe the implementation as follows~\cite{Nielsen02}. Given $x,y\in\{0,1\}^n$, we firstly compute $d=Ham(x,y)$, then choose a Gray code connecting $x$ and $y$. For instance, we choose the Gray code $(x,s_1,s_2,\cdots,s_{d-1},y)$ which has exactly $d+1$ elements. The basic idea is to perform a sequence of gates affecting the state changes $|x\rangle\rightarrow|s_1\rangle\rightarrow\cdots\rightarrow|s_{d-1}\rangle$, then to perform a controlled-$U$ operation($U$ is a $2\times2$ unitary matrix), with the target qubit located at the single bit where $s_{d-1}$ and $y$ differ, and then to undo the first stage, transforming $|s_{d-1}\rangle\rightarrow|s_{d-2}\rangle\rightarrow\cdots\rightarrow|s_1\rangle\rightarrow|x\rangle$. Note that if we intend to implement a near-trivial rotation, then $U=R(\theta)$, on the other hand ,if we intend to implement a near-trivial phase shift, then $U=\left(\begin{array}{cc}1&0\\0&e^{i\theta}\end{array}\right) $or $U=\left(\begin{array}{cc}e^{i\theta} &0 \\ 0 & 1\end{array}\right)$.

The above description could be implemented by a circuit with generalized Toffoli gates and $n$-qubit controlled-$U$ operation. We give an example to explain how to construct two-level unitary transformation. Suppose we intend to implement a Hadamard transformation on the bases $x=010$ and $y=101$. With the aid of a Gray code $(010, 011, 001, 101)$, the quantum circuit is constructed as follows:

%\vspace*{8mm}
%\begin{center}
%\label{fig1}
%\centerline{\psfig{figure=fig1.eps,width=5cm}}\end{center}
%\vspace*{-5mm} {\footnotesize {\bf Figure 1}\quad Implementation of two-level unitary transformation which performs a Hadamard transformation on the bases $x=010$ and $y=101$.}% \vspace*{8mm}

\begin{figure} [htbp]
%\vspace*{13pt}
\centerline{\epsfig{file=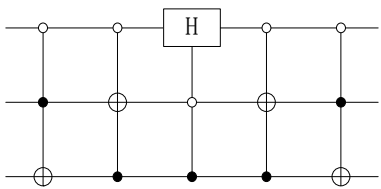, width=5cm}} %100 percent
\vspace*{5pt}
\caption{\label{fig1}Implementation of two-level unitary transformation which performs a Hadamard transformation on the bases $x=010$ and $y=101$.}
\end{figure}

Given $x,y$ and $U$( $U=R(\theta)$ or $U=P(\theta')$ ), we could construct a circuit to perform a near-trivial transformation($[x,y,\theta ,\theta']$) between dimensions $x$ and $y$. This implementation is not a universal but a uniformly construction, because we must construct different circuits to implement different near-trivial transformations $[x,y,\theta ,\theta']$ when any one of the arguments $x,y,\theta,\theta'$ is changed.

In Section~\ref{sec:uqc} we will give a universal construction for near-trivial transformation, in which the arguments $x,y,\theta,\theta'$ are inputs.

\subsection{Constructing a QTM to carry out near-trivial rotations}\label{sec:2.3}
We will briefly introduce the way in which Bernstein and Vazirani construct a QTM to carry out near-trivial rotations by angle $\theta$. The five steps are as follows~\cite{Bernstein97}:
\begin{enumerate}
  \item Calculate $k$ such that $k\Re$ mod $2\pi\in[\theta-\epsilon,\theta+\epsilon]$, where $\Re=2\pi\sum^\infty_{i=1}2^{-2^i}$.
  \item Transform $w,x,y$ into $b,x,y,z$, where $b=\left\{\begin{array}{r@{~,~}l}0 & w=x \\ 1 & w\neq x, w=y \\ \sharp & w\neq x, w\neq y \end{array}\right.$,$z=\left\{\begin{array}{r@{~,~}l}w&b=\sharp\\empty&else\end{array}\right.$.
  \item Run the rotation applying machine $k$ times on the first bit of $b$.
  \item Reverse step~2 transforming $\sharp,x,y,w$ with $w\neq x,y$ into $w,x,y$, transforming $0,x,y$ into $x,x,y$ and transforming $1,x,y$ with $x\neq y$ into $y,x,y$.
  \item Reverse step~1 erasing $k$.
\end{enumerate}

The desired QTM can be built by constructing a QTM for each of these five steps and then dovetailing them together.

\section{Universal quantum circuit for single-qubit transformations}
Arbitrary single-qubit unitary transformation can be written in the form~\cite{Nielsen02}
%\begin{equation}U=exp(i\alpha)R_{\hat{n}}(\theta)\end{equation}
\begin{equation}U=exp(i\alpha)R_z(\beta)R_y(\gamma)R_z(\delta),\label{eq2}\end{equation}
for some real numbers $\alpha$,$\beta$,$\gamma$ and $\delta$. In this section, we first construct two universal quantum circuits for single-qubit rotations $R_y(\theta)$ and $R_z(\theta)$, and then propose a universal quantum circuit implementing $U$.

In this paper, $\eta>0$ is a fixed angle and we can take $\eta=2\pi$. The value of $m$ is an interger determined by the required precision.

{\bf Lemma~1:} There exists a universal quantum circuit $C_1$ such that $C_1(|d\rangle\otimes|r_1r_2\cdots r_m\rangle)=(R_y(\theta)|d\rangle)\otimes|r_1r_2\cdots r_m\rangle$, where $|d\rangle$ is a single-qubit state, $r_i\in\{0,1\},\forall i\in\{1,2,\cdots,m\}$ and $\theta=(0.r_1r_2\cdots r_m)\cdot2\pi$.

{\bf Proof:}
For any $\theta\in[0,2\pi)$, we calculate the binary value of $\frac{\theta}{\eta}$. Denote the approximate value of $\frac{\theta}{\eta}$ as $0.r_1r_2\cdots r_m$, it can be deduced that $\theta\approx(0.r_1r_2\cdots r_m)\eta=(r_1\cdot2^{-1}+r_2\cdot2^{-2}+\cdots+r_m\cdot2^{-m})\eta=r_1\cdot\frac{\eta}{2}+r_2\cdot\frac{\eta}{2^2}+\cdots+r_m\cdot\frac{\eta}{2^m}$.
Since $R_y(\theta_1+\theta_2)=R_y(\theta_1)R_y(\theta_2)$~\cite{Barenco95}, $R_y(\theta)$ can be approximated with $R_y(r_1\cdot\frac{2\pi}{2})R_y(r_2\cdot\frac{2\pi}{2^2})\cdots R_y(r_m\cdot\frac{2\pi}{2^m})$, where $0.r_1r_2\cdots r_m$ is the $m$ decimal approximation of $\frac{\theta}{\eta}$.

We obtain a universal quantum circuit which can approximately implement unitary transformation $R_y(\theta)$ for any $\theta\in[0,2\pi)$. The $m$ variables $r_1r_2\cdots r_m$ are input parameters for this circuit. The circuit is as follows:

%\vspace*{8mm}
%\begin{center}
%\label{fig2}
%\centerline{\psfig{figure=fig2.eps,width=9cm}}\end{center}
%\vspace*{-5mm} {\footnotesize {\bf Figure 2}\quad The circuit on the right is an approximately implementation of the left one. The inputs $r_1r_2\cdots r_m$ are determined by the value of $\theta$ and satisfy the condition $0.r_1r_2\cdots r_m\approx\frac{\theta}{\eta}$, where $\eta$ is a fixed angle.}% \vspace*{8mm}

\begin{figure} [htbp]
%\vspace*{13pt}
\centerline{\epsfig{file=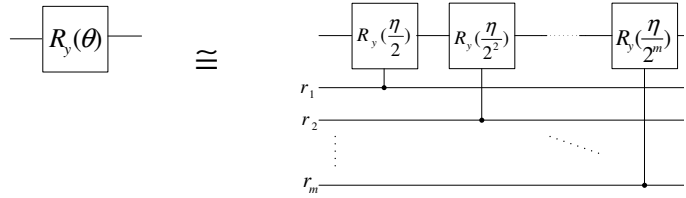, width=9cm}} %100 percent
\vspace*{5pt}
\caption{\label{fig2}The circuit on the right is an approximately implementation of the left one. The inputs $r_1r_2\cdots r_m$ are determined by the value of $\theta$ and satisfy the condition $0.r_1r_2\cdots r_m\approx\frac{\theta}{\eta}$, where $\eta$ is a fixed angle.}
\end{figure}

\begin{figure} [htbp]
%\vspace*{13pt}
\centerline{\epsfig{file=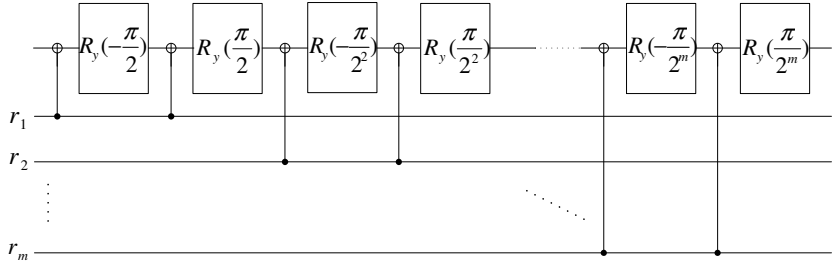, width=11cm}} %100 percent
\vspace*{5pt}
\caption{\label{fig3}Universal quantum circuit $C_1$ for $R_y(\theta)$. It consists of CNOT and other gates
in $\{R_y(\pm\frac{\pi}{2^j}), j\in{\bf N}\}$. In Figure~\ref{fig2}, we take $\eta=2\pi$ and replace all the controlled-$R_y(*)$ gates
with CNOT and single-qubit rotations. The inputs $r_1r_2\cdots r_m$ are the encoding of the single-qubit rotation $R_y(\theta)$.}
\end{figure}

Because of the identity $R_y(2\phi)=R_y(\phi)XR_y(-\phi)X$, the right circuit in Figure~\ref{fig2} is equivalent to the circuit $C_1$ in Figure~\ref{fig3}.

%%\vspace*{8mm}
%\begin{center}
%\label{fig3}
%\centerline{\psfig{figure=fig3.eps,width=11cm}}\end{center}
%\vspace*{-5mm} {\footnotesize {\bf Figure 3}\quad Universal quantum circuit $C_1$ for $R_y(\theta)$. It consists of CNOT and other gates
%in $\{R_y(\pm\frac{\pi}{2^j}), j\in{\bf N}\}$. In Figure 2, we take $\eta=2\pi$ and replace all the controlled-$R_y(*)$ gates
%with CNOT and single-qubit rotations. The inputs $r_1r_2\cdots r_m$ are the encoding of the single-qubit rotation $R_y(\theta)$.}% \vspace*{8mm}

Because $C_1(|d\rangle\otimes|r_1r_2\cdots r_m\rangle)=(R_y(\theta)|d\rangle)\otimes|r_1r_2\cdots r_m\rangle$ when $\theta=(0.r_1r_2\cdots r_m)\cdot2\pi$, the inputs $r_1r_2\cdots r_m$ for the circuit $C_1$ are the encoding of the single-qubit rotation $R_y(\theta)$ with $\frac{\theta}{2\pi}=0.r_1r_2\cdots r_m$.~$\Box$

%When $\theta\in(-2\pi,0)$, we calculate the binary value of $\frac{2\pi+\theta}{\eta}$ to be accurate to $m$ decimal places. If we denote this approximate value of $\frac{2\pi+\theta}{\eta}$ as $0.r_1r_2\cdots r_m$, then $2\pi+\theta\approx r_1\cdot\frac{\eta}{2}+r_2\cdot\frac{\eta}{2^2}+\cdots+r_m\cdot\frac{\eta}{2^m}$. Since $R_y(\theta)=R_y(2\pi+\theta)$ and $R_y(\theta_1+\theta_2)=R_y(\theta_1)R_y(\theta_2)$, $R_y(\theta)$ can be approximated using $R_y(r_1\cdot\frac{\eta}{2})R_y(r_2\cdot\frac{\eta}{2^2})\cdots R_y(r_m\cdot\frac{\eta}{2^m})$, where $-2\pi<\theta<0$ and $0.r_1r_2\cdots r_m$ is the $m$ decimal approximation of $\frac{2\pi+\theta}{\eta}$. Thus, we can

{\bf Lemma~2:} There exists a universal quantum circuit $C_2$ such that $C_2(|d\rangle\otimes|r_1r_2\cdots r_m\rangle)=(R_z(\theta)|d\rangle)\otimes|r_1r_2\cdots r_m\rangle$, where $|d\rangle$ is a single-qubit state, $r_i\in\{0,1\},\forall i\in\{1,2,\cdots,m\}$ and $\theta=(0.r_1r_2\cdots r_m)\cdot2\pi$.

{\bf Proof:}
Since $R_z(\theta_1+\theta_2)=R_z(\theta_1)R_z(\theta_2)$ and $R_z(2\phi)= R_z(\phi)XR_z(-\phi)X$ \cite{Barenco95}, $R_z(\theta)$ can be approximated using $R_z(r_1\cdot\frac{\eta}{2})R_z(r_2\cdot\frac{\eta}{2^2})\cdots R_z(r_m\cdot\frac{\eta}{2^m})$, where $\theta\in[0,2\pi)$ and $0.r_1r_2\cdots r_m$ is the $m$ decimal approximation of $\frac{\theta}{\eta}$.

\begin{figure} [htbp]
%\vspace*{13pt}
\centerline{\epsfig{file=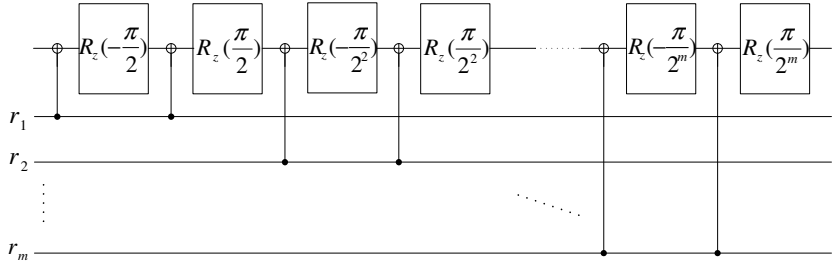, width=11cm}} %100 percent
\vspace*{5pt}
\caption{\label{fig4}Universal quantum circuit $C_2$ for $R_z(\theta)$. It consists of CNOT and the gates in $\{R_z(\pm\frac{\pi}{2^j}), j\in{\bf N}\}$. The inputs $r_1r_2\cdots r_m$ are the encoding of the single-qubit rotation $R_z(\theta)$.}
\end{figure}

In the same way as the construction in Lemma~1, we construct a universal quantum circuit $C_2$ in Figure~\ref{fig4} which can approximately implement unitary transformation $R_z(\theta)$ for any $\theta\in[0,2\pi)$. Because $C_2(|d\rangle\otimes|r_1r_2\cdots r_m\rangle)=(R_z(\theta)|d\rangle)\otimes|r_1r_2\cdots r_m\rangle$ when $\theta=(0.r_1r_2\cdots r_m)\cdot2\pi$, the inputs $r_1r_2\cdots r_m$ for the circuit $C_2$ are the encoding of the single-qubit rotation $R_z(\theta)$ with $\frac{\theta}{2\pi}=0.r_1r_2\cdots r_m$.~$\Box$

%%\vspace*{8mm}
%\begin{center}
%\label{fig4}
%\centerline{\psfig{figure=fig4.eps,width=11cm}}\end{center}
%\vspace*{-5mm} {\footnotesize {\bf Figure 4}\quad Universal quantum circuit $C_2$ for $R_z(\theta)$. It consists of CNOT and the gates in $\{R_z(\pm\frac{\pi}{2^j}), j\in{\bf N}\}$. The inputs $r_1r_2\cdots r_m$ are the encoding of the single-qubit rotation $R_z(\theta)$.}% \vspace*{8mm}

{\bf Theorem~1:} There exists a universal quantum circuit $C$ approximately implementing arbitrary single-qubit unitary transformation $U$. The circuit $C$ consists of CNOT and the gates in $\{R_z(\pm\frac{\pi}{2^j}), R_y(\pm\frac{\pi}{2^j}), j\in{\bf N}\}$.

{\bf Proof:} According to Lemma~1 and Lemma~2, we can construct universal quantum circuits $C_1$ and $C_2$ implementing $R_y(\gamma)$ and $R_z(\delta)$ respectively. Moreover, because of ''Eq.~(\ref{eq2})'', the universal circuit $C$ is accomplished by connecting one $C_1$ and two $C_2$. The connecting order is $C_2,C_1,C_2$. If $\frac{\delta}{2\pi}\approx0.e_1,\frac{\gamma}{2\pi}\approx0.e_2,\frac{\beta}{2\pi}\approx0.e_3$, then the encoding for the single-qubit unitary transformation is $e_1e_2e_3$, where $e_1$ is the encoding of $R_z(\delta)$ in $C_2$, $e_2$ is the encoding of $R_y(\gamma)$ in $C_1$ and $e_3$ is the encoding of $R_z(\beta)$ in $C_2$. Thus, we have the equation $C(|d\rangle|e_1\rangle|e_2\rangle|e_3\rangle)=(U'|d\rangle)|e_1\rangle|e_2\rangle|e_3\rangle$, where $U'=R_z(0.e_3\cdot2\pi)R_y(0.e_2\cdot2\pi)R_z(0.e_1\cdot2\pi)\approx R_z(\beta)R_y(\gamma)R_z(\delta)$.~$\Box$
\section{Universal quantum circuit for near-trivial transformations}\label{sec:uqc}
\subsection{Overall steps}\label{sec:3.1}
\noindent
We modify the five steps in Section~\ref{sec:2.3}. Then any near-trivial transformation can be realized by the following 3 steps:

We use $4$ quantum registers denoted as $w,x,y,b$. The registers $w,x$ and $y$ are $n$-qubit quantum registers which are used to store inputs, and $b$ is a $2$-qubit register initiated with quantum state $|10\rangle$. In the following description, for convenience, we still use $w,x,y,b$ to represent the contents of the $4$ quantum registers.
\begin{description}
  \item[a] Transform $w,x,y,b$ into $z,x,y,b'$, where $b'=\left\{\begin{array}{r@{,}l}00&w=x,w\neq y \\ 01 & w\neq x,w=y \\ 10 & w\neq x,w\neq y \\ 11 & w=x,w=y \end{array}\right.$, $z=\left\{\begin{array}{r@{,}l}0 & b'\in\{00,01\} \\ w & b'\in\{10,11\}\end{array}\right.$.
  \item[b] Perform unitary transformation $R(\theta)=\left(\begin{array}{cc}cos\theta & -sin\theta \\ sin\theta & cos\theta \end{array}\right)$ on the second qubit of $b'$ if $b'=00$ or $b'=01$. Perform unitary transformation $P(\theta')=\left(\begin{array}{cc}1&0\\0&e^{i\theta'}\end{array}\right)$ on the second qubit of $b'$ if $b'=10$ or $b'=11$.
  \item[c] Reverse step~{\bf a}.
\end{description}

The above three transformations are all unitary and are denoted as $U_a$, $U_b$, $U_c$ respectively. It is oblivious that $U_c=U^{-1}_a$, so $U_c$ can be implemented by the mirror image of the quantum circuit implementing $U_a$~\cite{Fredkin82,Toffoli80,Toffoli81,Bennett73}. Therefore, we only need to construct quantum circuits for $U_a$ and $U_b$. Then, by connecting the corresponding qubits of the three circuits in the order $U_a, U_b, U_c$, we obtain a quantum circuit implementing near-trivial transformations $[x,y,\theta ,\theta']$, where $x,y\in\{0,1\}^n$, and $\theta,\theta'\in(0,2\pi)$. The circuit obtained is universal, because the parameters $x,y,\theta ,\theta'$ act as inputs for the circuit.

\subsection{Quantum circuit implementing $U_a$}\label{Ua}
\noindent
We will take three steps to construct a quantum circuit implementing $U_a$. Suppose the four quantum registers are initiated with $|w\rangle|x\rangle|y\rangle|b\rangle$.

In the first step, $b'$ is obtained after performing the unitary operation $U_1$. The transformation $U_1$ is as follows:
\begin{equation}U_1|w\rangle|x\rangle|y\rangle|b\rangle=|w\rangle|x\oplus w\rangle|y\oplus w\rangle|b'\rangle.\end{equation}
In this transformation, the first quantum register remains the same and we obtain the required value of $b'$ in the fourth quantum register. In addition, the second and third quantum registers are changed. However, we can restore the states of the two quantum registers in the second step. The second transformation $U_2$ is as follows:
\begin{equation}U_2|w\rangle|x\oplus w\rangle|y\oplus w\rangle|b'\rangle=|w\rangle|x\rangle|y\rangle|b'\rangle.\end{equation}
Through the above two steps, the value of $b'$ is obtained in the fourth quantum register while keeping the values of $w,x,y$ unchanged. In the third step, we can obtain the required value of $z$ from $b'$ and $w$. Since $z=w$ when $b'\in\{10,11\}$, the first quantum register should keep unchanged when $b'$ equals to $10$ or $11$. If $b'$ equals to $00$ or $01$, we could infer that the contents of the first register is the same as the second or the third respectively($w=x$ or $w=y$). So the first quantum register can be changed into $0$ if $b'\in\{00,01\}$. The third transformation $U_3$ is as follows:
\begin{equation}U_3|w\rangle|x\rangle|y\rangle|b'\rangle=|z\rangle|x\rangle|y\rangle|b'\rangle.\end{equation}

\begin{figure} [htbp]
%\vspace*{13pt}
\centerline{\epsfig{file=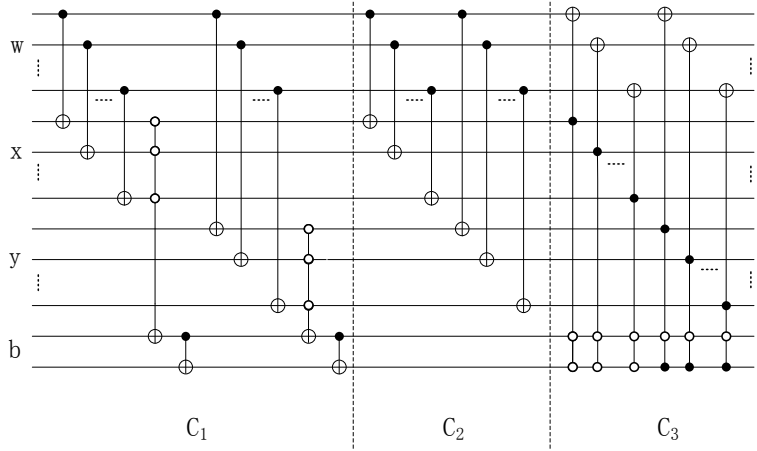, width=10cm}} %100 percent
\vspace*{5pt}
\caption{\label{fig5}Quantum circuit $C_a$ implementing the transformation $U_a$.
The registers $w,x$ and $y$ are $n$-qubit quantum registers which are used to store inputs,
and $b$ is a $2$-qubit register initiated with quantum state $|10\rangle$.
This quantum circuit is divided into three parts denoted as $C_1$,$C_2$ and $C_3$
which implement $U_1$, $U_2$ and $U_3$ respectively.}
\end{figure}

From the above analysis, we construct the quantum circuit in Figure~\ref{fig5} implementing $U_a$(the whole quantum circuit is divided into three parts denoted as $C_1$,$C_2$ and $C_3$ which implement the above three transformations respectively):

%%\vspace*{8mm}
%\begin{center}
%\label{fig5}
%\centerline{\psfig{figure=fig5.eps,width=10cm}}\end{center}
%\vspace*{-5mm} {\footnotesize {\bf Figure 5}\quad Quantum circuit $C_a$ implementing the transformation $U_a$.
%The registers $w,x$ and $y$ are $n$-qubit quantum registers which are used to store inputs,
%and $b$ is a $2$-qubit register initiated with quantum state $|10\rangle$.
%This quantum circuit is divided into three parts denoted as $C_1$,$C_2$ and $C_3$
%which implement $U_1$, $U_2$ and $U_3$ respectively.}% \vspace*{8mm}

{\bf Proposition~1:} The quantum circuit $C_a$ can implement the transformation $U_a$ stated in Section~\ref{sec:3.1}~a.

\vspace*{12pt}
\noindent
{\bf Proof:}
For implicity, we denote $C_a$ as the transformation implemented by the quantum circuit $C_a$ in Figure~\ref{fig5}.
\begin{description}
  \item[(1)] If $w=x$ and $w\neq y$, $C_a|w\rangle|x\rangle|y\rangle|b\rangle=C_3C_2C_1|x\rangle|x\rangle|y\rangle|10\rangle\\=C_3C_2|x\rangle|0\rangle|y\oplus x\rangle|00\rangle=C_3|x\rangle|x\rangle|y\rangle|00\rangle=|0\rangle|x\rangle|y\rangle|00\rangle$.
  \item[(2)] If $w\neq x$ and $w=y$,  $C_a|w\rangle|x\rangle|y\rangle|b\rangle=C_3C_2C_1|y\rangle|x\rangle|y\rangle|10\rangle\\=C_3C_2|y\rangle|x\oplus y\rangle|0\rangle|01\rangle=C_3|y\rangle|x\rangle|y\rangle|01\rangle=|0\rangle|x\rangle|y\rangle|01\rangle$.
  \item[(3)] If $w\neq x$ and $w\neq y$,  $C_a|w\rangle|x\rangle|y\rangle|b\rangle=C_3C_2C_1|w\rangle|x\rangle|y\rangle|10\rangle\\=C_3C_2|w\rangle|x\oplus w\rangle|y\oplus w\rangle|10\rangle=C_3|w\rangle|x\rangle|y\rangle|10\rangle=|w\rangle|x\rangle|y\rangle|10\rangle$.
  \item[(4)] If $w=x$ and $w=y$,  $C_a|w\rangle|x\rangle|y\rangle|b\rangle=C_3C_2C_1|x\rangle|x\rangle|x\rangle|10\rangle\\=C_3C_2|x\rangle|0\rangle|0\rangle|11\rangle =C_3|x\rangle|x\rangle|x\rangle|11\rangle=|x\rangle|x\rangle|x\rangle|11\rangle=|w\rangle|x\rangle|y\rangle|11\rangle$.
\end{description}
From {\bf(1)},{\bf(2)},{\bf(3)} and {\bf(4)}, we know that $C_a|w\rangle|x\rangle|y\rangle|b\rangle=|z\rangle|x\rangle|y\rangle|b'\rangle$. Thus $C_a$ implements the transformation $U_a$.~$\Box$

Because the circuit implementing $U_c$ is the mirror image of the circuit $C_a$(The mirror image circuit of $C_a$ is denoted as $C_a^{-1}$), the transformation $U_c$ can be represented as follows:

$U_c|0\rangle|x\rangle|y\rangle|00\rangle=|x\rangle|x\rangle|y\rangle|10\rangle$, if $x,y\in\{0,1\}^n$ and $x\neq y$;

$U_c|0\rangle|x\rangle|y\rangle|01\rangle=|y\rangle|x\rangle|y\rangle|10\rangle$, if $x,y\in\{0,1\}^n$ and $x\neq y$;

$U_c|w\rangle|x\rangle|y\rangle|10\rangle=|w\rangle|x\rangle|y\rangle|10\rangle$, if $w,x,y\in\{0,1\}^n$ and $w\neq x, w\neq y$;

$U_c|x\rangle|x\rangle|x\rangle|11\rangle=|x\rangle|x\rangle|x\rangle|10\rangle$, $\forall x\in\{0,1\}^n$.

\subsection{Quantum circuit implementing $U_b$}\label{Ub}
\begin{figure} [htbp]
%\vspace*{13pt}
\centerline{\epsfig{file=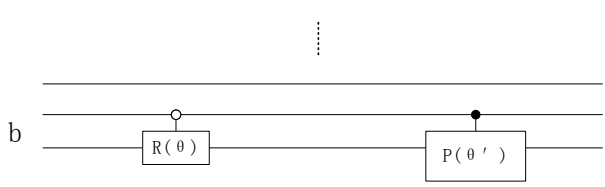, width=8cm}} %100 percent
\vspace*{5pt}
\caption{\label{fig6}Quantum circuit $C_b$ implementing the transformation $U_b$. In this quantum circuit, the two controlled unitary transformation act on the $4$th quantum register $b$. It carries out a trivial transformation on the quantum registers $w,x,y$.}
\end{figure}

\noindent
The transformation $U_b$ only affects the two qubits of the register $b$, and the first qubit of $b$ is control qubit and the other is target qubit. We can easily construct the quantum circuit in Figure~\ref{fig6} implementing $U_b$.

%%\vspace*{8mm}
%\begin{center}
%\label{fig6}
%\centerline{\psfig{figure=fig6.eps,width=8cm}}\end{center}
%\vspace*{-5mm} {\footnotesize {\bf Figure 6}\quad Quantum circuit $C_b$ implementing the transformation $U_b$. In this quantum circuit, the two controlled unitary transformation act on the $4$th quantum register $b$. It carries out a trivial transformation on the quantum registers $w,x,y$.}% \vspace*{8mm}

Next, we will prove that the quantum circuit $C_b$ can accomplish the transformation $U_b$.

{\bf Proposition~2:} The quantum circuit $C_b$ can implement the transformation $U_b$ stated in Section~\ref{sec:3.1}~b.

\vspace*{12pt}
\noindent
{\bf Proof:}
After performing the transformation $U_a$, the register $b$ may be in the superposition of the $4$ quantum states $|00\rangle,|01\rangle,|10\rangle,|11\rangle$.
\begin{description}
  \item[(1)] If $b'=00$, $C_b|z\rangle|x\rangle|y\rangle|00\rangle=|z\rangle|x\rangle|y\rangle C^1_0(R(\theta))|00\rangle=|z\rangle|x\rangle|y\rangle|0\rangle R(\theta)|0\rangle\\=|z\rangle|x\rangle|y\rangle|0\rangle(cos\theta|0\rangle+sin\theta|1\rangle)$.
  \item[(2)] If $b'=01$, $C_b|z\rangle|x\rangle|y\rangle|01\rangle=|z\rangle|x\rangle|y\rangle C^1_0(R(\theta))|01\rangle=|z\rangle|x\rangle|y\rangle|0\rangle R(\theta)|1\rangle\\=|z\rangle|x\rangle|y\rangle|0\rangle(-sin\theta|0\rangle+cos\theta|1\rangle)$.
  \item[(3)] If $b'=10$, $C_b|z\rangle|x\rangle|y\rangle|10\rangle=|z\rangle|x\rangle|y\rangle C(P(\theta'))|10\rangle=|z\rangle|x\rangle|y\rangle|1\rangle P(\theta')|0\rangle\\=|z\rangle|x\rangle|y\rangle|10\rangle$.
  \item[(4)] If $b'=11$, $C_b|z\rangle|x\rangle|y\rangle|11\rangle=|z\rangle|x\rangle|y\rangle C(P(\theta'))|11\rangle=|z\rangle|x\rangle|y\rangle|1\rangle P(\theta')|1\rangle\\=|z\rangle|x\rangle|y\rangle e^{i\theta'}|11\rangle$.
\end{description}
From {\bf(1)},{\bf(2)},{\bf(3)} and {\bf(4)}, the verification is completed.~$\Box$

\begin{figure} [htbp]
%\vspace*{13pt}
\centerline{\epsfig{file=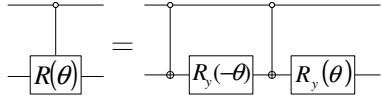, width=5cm}} %100 percent
\vspace*{5pt}
\caption{\label{fig7}Implementing $C^1_0(R(\theta))$ using generalized CNOT and single-qubit operations.}
\end{figure}

\begin{figure} [htbp]
%\vspace*{13pt}
\centerline{\epsfig{file=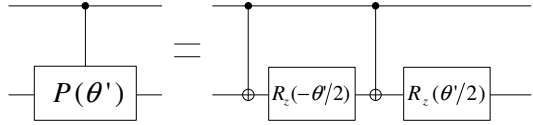, width=7cm}} %100 percent
\vspace*{5pt}
\caption{\label{fig8}Implementing $C(P(\theta'))$ using CNOT and single-qubit operations.}
\end{figure}

Furthermore, we will show that $U_b$ can be implemented with a quantum circuit consists of only CNOT and single-qubit rotations. Because of the identities $R(\theta)=R_y(\theta)XR_y(-\theta)X$ and $P(\theta')=e^{i\theta'/2}R_z(\frac{\theta'}{2})XR_z(-\frac{\theta'}{2})X$~\cite{Barenco95}, we have the following two circuit identities in Figure~\ref{fig7} and Figure~\ref{fig8}.

%%\vspace*{4mm}
%\begin{center}
%\label{fig7}
%\centerline{\psfig{figure=fig7.eps,width=5cm}}\end{center}
%\vspace*{-5mm} {\footnotesize {\bf Figure 7}\quad Implementing $C^1_0(R(\theta))$ using generalized CNOT and single-qubit operations.}% \vspace*{8mm}

%\vspace*{4mm}
%\begin{center}
%\label{fig8}
%\centerline{\psfig{figure=fig8.eps,width=7cm}}\end{center}
%\vspace*{-5mm} {\footnotesize {\bf Figure 8}\quad Implementing $C(P(\theta'))$ using CNOT and single-qubit operations.}% \vspace*{8mm}

{\bf Theorem~2:} There exists universal quantum circuit $C_1$ for $R_y(\theta)$ and $C_1'$ for $R_y(-\theta)$, and the encoding of $R_y(\theta)$ in $C_1$ is the same as the encoding of $R_y(-\theta)$ in $C_1'$, $\forall \theta\in[0,2\pi)$. In other words, there exists two quantum circuits $C_1$ and $C_1'$, and $r_1r_2\cdots r_m\in\{0,1\}^m$, such that $C_1(|d\rangle\otimes|r_1r_2\cdots r_m\rangle)=(R_y(\theta)|d\rangle)\otimes|r_1r_2\cdots r_m\rangle$ and $C_1'(|d\rangle\otimes|r_1r_2\cdots r_m\rangle)=(R_y(-\theta)|d\rangle)\otimes|r_1r_2\cdots r_m\rangle$, where $|d\rangle$ is a single-qubit state and $\theta=(0.r_1r_2\cdots r_m)\cdot 2\pi$.

{\bf Proof:} In Lemma~1, we have constructed a universal quantum circuit $C_1$ for $R_y(\theta)$,$\forall \theta\in[0,2\pi)$, and the encoding of $R_y(\theta)$ is $r_1r_2\cdots r_m$ where $0.r_1r_2\cdots r_m$ is the $m$ decimal approximation of $\frac{\theta}{2\pi}$.

\begin{figure} [htbp]
%\vspace*{13pt}
\centerline{\epsfig{file=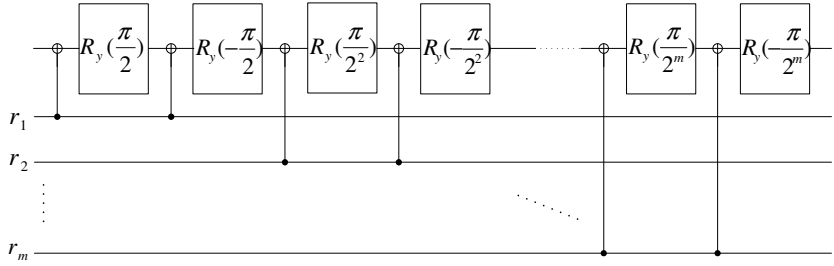, width=11cm}} %100 percent
\vspace*{5pt}
\caption{\label{fig9}Universal quantum circuit $C_1'$ for $R_y(-\theta)$. It consists of CNOT and the gates in $\{R_y(\pm\frac{\pi}{2^j}), j\in{\bf N}\}$. The inputs $r_1r_2\cdots r_m$, which satisfy $\frac{\theta}{\eta}\approx 0.r_1r_2\cdots r_m$, are the encoding of the single-qubit rotation $R_y(-\theta),\theta\in[0,2\pi)$.}
\end{figure}

Because $\frac{\theta}{\eta}\approx 0.r_1r_2\cdots r_m$, $-\theta\approx r_1\cdot\frac{-\eta}{2}+r_2\cdot\frac{-\eta}{2^2}+\cdots+r_m\cdot\frac{-\eta}{2^m}$. Thus $R_y(-\theta)$ can be approximated with $R_y(r_1\cdot\frac{-\eta}{2})R_y(r_2\cdot\frac{-\eta}{2^2})\cdots R_y(r_m\cdot\frac{-\eta}{2^m})$, where $0\leq\theta<2\pi$ and $0.r_1r_2\cdots r_m$ is the $m$ decimal approximation of $\frac{\theta}{\eta}$. In the same way as the construction of $R_y(\theta)$, we obtain a universal quantum circuit $C_1'$(in Figure~\ref{fig9}) for $R_y(-\theta)$, and the encoding of $R_y(-\theta)$ in $C_1'$ is $r_1r_2\cdots r_m$.

%\vspace*{8mm}
%\begin{center}
%\label{fig9}
%\centerline{\psfig{figure=fig9.eps,width=11cm}}\end{center}
%\vspace*{-5mm} {\footnotesize {\bf Figure 9}\quad Universal quantum circuit $C_1'$ for $R_y(-\theta)$. It consists of CNOT and the gates in $\{R_y(\pm\frac{\pi}{2^j}), j\in{\bf N}\}$. The inputs $r_1r_2\cdots r_m$, which satisfy $\frac{\theta}{\eta}\approx 0.r_1r_2\cdots r_m$, are the encoding of the single-qubit rotation $R_y(-\theta),\theta\in[0,2\pi)$.}% \vspace*{8mm}

Thus, the proof is completed.~$\Box$

{\bf Corollary~1:} $C^1_0(R(\theta))$ can be implemented by a universal quantum circuit $C_R$ which consists of only CNOT and the gates in $\{R_y(\pm\frac{\pi}{2^j}), j\in{\bf N}\}$, and the encoding of $C^1_0(R(\theta))$ is the same as the encoding of $R_y(\theta)$ in $C_1$.

{\bf Proof:} On the right side of the circuit in Figure~\ref{fig7}, we replace the single-qubit rotations $R_y(\theta)$ and $R_y(-\theta)$ with $C_1$ and $C_1'$, respectively. According to Theorem~2, since the encoding of $R_y(\theta)$ in $C_1$ is the same as the encoding of $R_y(-\theta)$ in $C_1'$, we can connect the tail of encoding wire in $C_1$ with the head of encoding wire in $C_2$. Then, we will get the required quantum circuit $C_R$.~$\Box$

We can come up with a similar result in the following:

{\bf Theorem~3:} There exists universal quantum circuit $C_2$ for $R_z(\theta)$ and $C_2'$ for $R_z(-\theta)$, and the encoding of $R_z(\theta)$ in $C_2$ is the same as the encoding of $R_z(-\theta)$ in $C_2'$, $\forall \theta\in[0,2\pi)$.

\begin{figure} [htbp]
%\vspace*{13pt}
\centerline{\epsfig{file=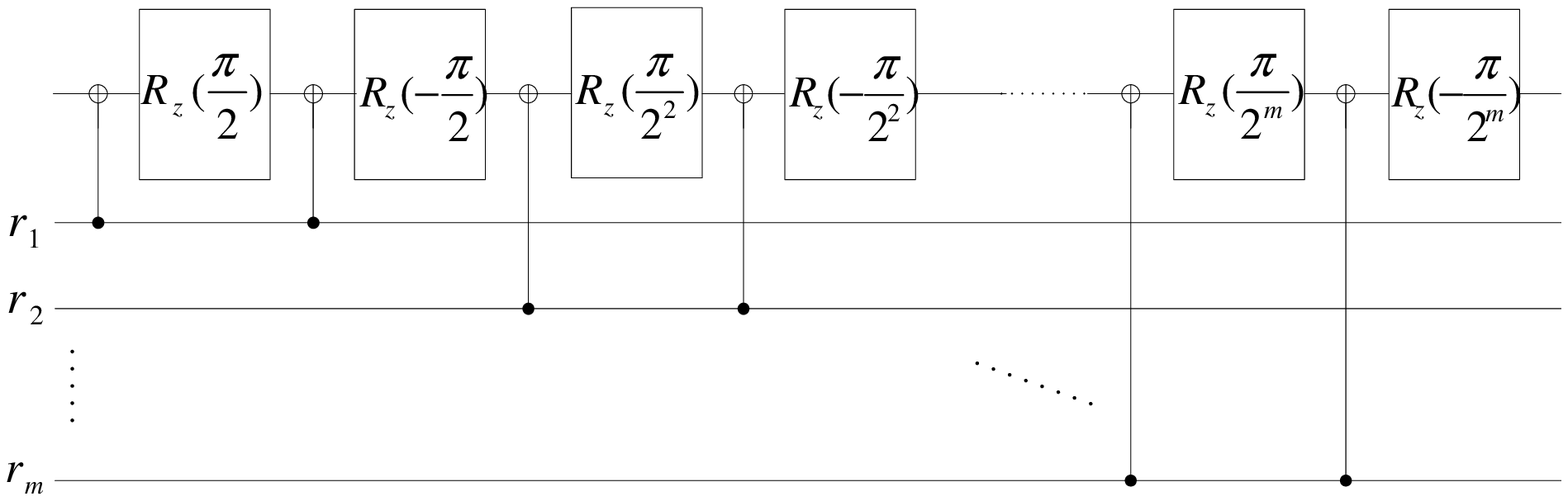, width=11cm}} %100 percent
\vspace*{5pt}
\caption{\label{fig10}Universal quantum circuit $C_2'$ for $R_z(-\theta)$. It consists of CNOT and the gates in $\{R_z(\pm\frac{\pi}{2^j}), j\in{\bf N}\}$. The inputs $r_1r_2\cdots r_m$, which satisfy $\frac{\theta}{\eta}\approx 0.r_1r_2\cdots r_m$, are the encoding of the single-qubit rotation $R_z(-\theta),\theta\in[0,2\pi)$.}
\end{figure}

{\bf Proof:} The circuit $C_2$ is constructed in Lemma~2. In a similar way, we provide the universal quantum circuit $C_2'$ for $R_z(-\theta)$ in Figure~\ref{fig10}.~$\Box$
%\vspace*{8mm}
%\begin{center}
%\label{fig10}
%\centerline{\psfig{figure=fig10.eps,width=11cm}}\end{center}
%\vspace*{-5mm} {\footnotesize {\bf Figure 10}\quad Universal quantum circuit $C_2'$ for $R_z(-\theta)$. It consists of CNOT and the gates in $\{R_z(\pm\frac{\pi}{2^j}), j\in{\bf N}\}$. The inputs $r_1r_2\cdots r_m$, which satisfy $\frac{\theta}{\eta}\approx 0.r_1r_2\cdots r_m$, are the encoding of the single-qubit rotation $R_z(-\theta),\theta\in[0,2\pi)$.}% \vspace*{8mm}

{\bf Corollary~2:} $C(P(\theta'))$ can be implemented by a universal quantum circuit $C_P$ which consists of only CNOT and the gates in $\{R_z(\pm\frac{\pi}{2^j}), j\in{\bf N}\}$, and the encoding of $C(P(\theta'))$ is the same as the encoding of $R_z(\frac{\theta'}{2})$ in $C_2$.

{\bf Proof:} The construction of $C_P$ is the same as the construction of $C_R$ in Corrollary~1. On the right side of the circuit in Figure~\ref{fig8}, we obtain the circuit $C_P$ by replacing the single-qubit rotations $R_z(\frac{\theta'}{2})$ and $R_z(-\frac{\theta'}{2})$ with $C_2$ and $C_2'$, respectively. If $\frac{\theta'/2}{2\pi}=0.r_1r_2\cdots r_m$, the encoding of $R_z(\frac{\theta'}{2})$ in $C_2$ is $r_1r_2\cdots r_m$, and so is the encoding of $R_z(\frac{\theta'}{2})$ in $C_2$. Thus, $C_P(|b\rangle\otimes|r_1r_2\cdots r_m\rangle)=(C(P(\theta'))|b\rangle)\otimes|r_1r_2\cdots r_m\rangle$, where $\theta'=(0.r_1r_2\cdots r_m)\cdot 4\pi$.~$\Box$

\begin{figure} [htbp]
%\vspace*{13pt}
\centerline{\epsfig{file=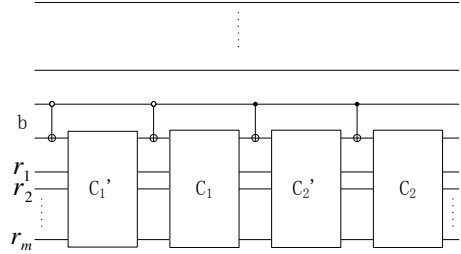, width=6cm}} %100 percent
\vspace*{5pt}
\caption{\label{fig11}Quantum circuit $C_b'$ which implements $U_b$ approximately. If the $1$st qubit of $b$ is $0$, the circuit would perform $C^1_0(R(\theta))$, where $\theta\approx(0.r_1r_2\cdots r_m)\cdot2\pi$. If the $1$st qubit of $b$ is $1$, then the circuit would perform $C(P(\theta'))$, where $\theta'\approx(0.r_1r_2\cdots r_m)\cdot4\pi$.}
\end{figure}

{\bf Proposition~3:} By connecting the universal quantum circuits for $C^1_0(R(\theta))$ and $C(P(\theta'))$, we can obtain a universal quantum circuit $C_b'$(in Figure~\ref{fig11}) approximately implementing $U_b$.

{\bf Proof:} In Figure~\ref{fig11}, quantum state $|b\rangle$ has two qubits. If the first qubit of $b$ is $0$, it can be deduced that $C_b'(|w,x,y\rangle|b\rangle|r_1r_2\cdots r_m\rangle)\\=|w,x,y\rangle C_R(|b\rangle|r_1r_2\cdots r_m\rangle)=|w,x,y\rangle(C^1_0(R(\theta))|b\rangle)|r_1r_2\cdots r_m\rangle$, where $\frac{\theta}{2\pi}=0.r_1r_2\cdots r_m$. If the first qubit of $b$ is $1$, $C_b'(|w,x,y\rangle|b\rangle|r_1r_2\cdots r_m\rangle)\\=|w,x,y\rangle C_P(|b\rangle|r_1r_2\cdots r_m\rangle)=|w,x,y\rangle(C(P(\theta'))|b\rangle)|r_1r_2\cdots r_m\rangle$, where $\frac{\theta'}{4\pi}=0.r_1r_2\cdots r_m$. Thus, the circuit $C_b'$ is an approximation of the circuit $C_b$ in Figure~\ref{fig6}.~$\Box$

%\vspace*{-10mm}
%\begin{center}
%\label{fig11}
%\centerline{\psfig{figure=fig11.eps,width=5cm}}\end{center}
%\vspace*{-8mm} {\footnotesize {\bf Figure 11}\quad Quantum circuit $C_b'$ which implements $U_b$ approximately. If the $1$st qubit of $b$ is $0$, the circuit would perform $C^1_0(R(\theta))$, where $\theta\approx(0.r_1r_2\cdots r_m)\cdot2\pi$. If the $1$st qubit of $b$ is $1$, then the circuit would perform $C(P(\theta'))$, where $\theta'\approx(0.r_1r_2\cdots r_m)\cdot4\pi$.}% \vspace*{8mm}
%~$\Box$

%It is worth to mention that:
{\bf Remark~1:}

(a) If we intend to perform controlled-$R(\theta)$(for any $\theta\in[0,2\pi)$) with $R(\theta)$ being performed on the $2$nd qubit of the register $b$, conditional on the $1$st qubit being set to $0$, then we take $r_1,r_2,\cdots,r_m$ as ancillary inputs for $C_b'$ where $r_1,r_2,\cdots,r_m$ are from the $m$ decimal approximation of $\frac{\theta}{\eta}$($\frac{\theta}{2\pi}$).

(b) If we intend to perform controlled-$P(\theta')$ (for any $\theta'\in[0,2\pi)$) with $P(\theta')$ being performed on the $2$nd qubit, conditional on the $1$st qubit being set to $1$, then we take $r_1,r_2,\cdots,r_m$ as ancillary inputs for $C_b'$ where $r_1,r_2,\cdots,r_m$ are from the $m$ decimal approximation of $\frac{\theta'/2}{\eta}$($\frac{\theta'}{4\pi}$).

\subsection{Universal quantum circuit for near-trivial transformations}
\begin{figure} [htbp]
%\vspace*{13pt}
\centerline{\epsfig{file=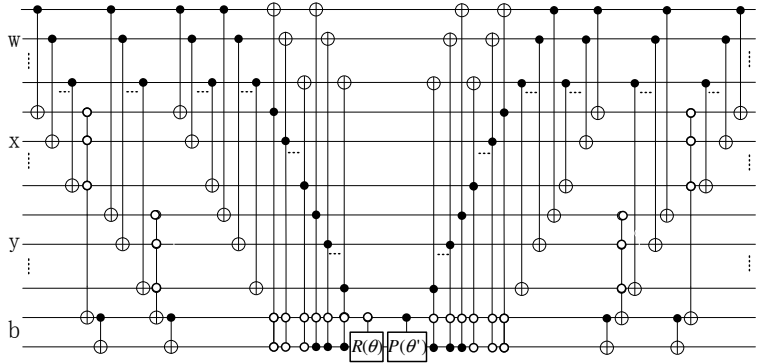, width=10cm}} %100 percent
\vspace*{5pt}
\caption{\label{fig12}Universal quantum circuit $C_U$ exactly implementing near-trivial transformations. We dovetail the circuits in Figure~\ref{fig5} and Figure~\ref{fig6} and the mirror image of the circuit in Figure~\ref{fig5}, then we obtain the whole circuit. In this circuit, there are two arguments $\theta$ and $\theta'$ which must be given previously. So, this circuit is not truly universal but partially universal quantum circuit implementing near-trivial transformations.}
\end{figure}

\noindent
So far, we have constructed quantum circuits $C_a$ and $C_b$ which implement $U_a$ and $U_b$, respectively. In addition, $U_c$ could be implemented with the mirror image of the circuit $C_a$. So, we obtain the quantum circuit $C_U$ (in Figure~\ref{fig12}) by connecting the corresponding qubits of the three circuits in order $C_a$, $C_b$, $C_a^{-1}$.

%\vspace*{-5mm}
%\begin{center}
%\label{fig12}
%\centerline{\psfig{figure=fig12.eps,width=10cm}}\end{center}
%\vspace*{-8mm} {\footnotesize {\bf Figure 12}\quad Universal quantum circuit $C_U$ exactly implementing near-trivial transformations. We dovetail the circuits in Figure 5 and Figure 6 and the mirror image of the circuit in Figure 5, then we obtain the whole circuit. In this circuit, there are two arguments $\theta$ and $\theta'$ which must be given previously. So, this circuit is not truly universal but partially universal quantum circuit implementing near-trivial transformations.}% \vspace*{8mm}

Note that the quantum circuit is not entirely universal because there are two arguments ($\theta$ and $\theta'$) in the quantum circuit. Here, for convenience, we give a verification for this quantum circuit. In the last part of this section, we will give an entirely universal quantum circuit which is an approximation of this circuit.

Next, we give a verification that the quantum circuit could implement arbitrary near-trivial transformation $[x,y,\theta,\theta']$, for any $x,y\in\{0,1\}^n$,where $\theta$ and $\theta'$ are certain fixed angles.

{\bf Proposition~4:} $C_U(|w\rangle\otimes|x,y,b\rangle)=([x,y,\theta,\theta']|w\rangle)\otimes|x,y,b\rangle$, for any $w,x,y\in\{0,1\}^n$, where $b=10$ and $\theta,\theta'\in[0,2\pi)$ are certain fixed angles.

{\bf Proof:}
In Proposition~1 and Proposition~2, we have analyzed the quantum circuits $C_a$, $C_b$ and $C_a^{-1}$ which implement $U_a$, $U_b$ and $U_c$, respectively.

Because $C_U$ is constructed by connecting the corresponding qubits of the three circuits in the order $C_a$, $C_b$, $C_a^{-1}$, we have $C_U=C_a^{-1}C_bC_a$.

\begin{description}
  \item[(1)] If $w=x$ and $w\neq y$,  $C_U|w\rangle|x\rangle|y\rangle|b\rangle=C_a^{-1}C_bC_a|x\rangle|x\rangle|y\rangle|10\rangle\\=C_a^{-1}C_b|0\rangle|x\rangle|y\rangle|00\rangle=C_a^{-1}|0\rangle|x\rangle|y\rangle|0\rangle(cos\theta|0\rangle+sin\theta|1\rangle)\\=(cos\theta|x\rangle+sin\theta|y\rangle)|x\rangle|y\rangle|10\rangle$.
  \item[(2)] If $w\neq x$ and $w=y$, $C_U|w\rangle|x\rangle|y\rangle|b\rangle=C_a^{-1}C_bC_a|y\rangle|x\rangle|y\rangle|10\rangle\\=C_a^{-1}C_b|0\rangle|x\rangle|y\rangle|01\rangle=C_a^{-1}|0\rangle|x\rangle|y\rangle|0\rangle(-sin\theta|0\rangle+cos\theta|1\rangle)\\=(-sin\theta|x\rangle+cos\theta|y\rangle)|x\rangle|y\rangle|10\rangle$.
  \item[(3)] If $w\neq x$ and $w\neq y$, $C_U|w\rangle|x\rangle|y\rangle|b\rangle=C_a^{-1}C_bC_a|w\rangle|x\rangle|y\rangle|10\rangle\\=C_a^{-1}C_b|w\rangle|x\rangle|y\rangle|10\rangle=C_a^{-1}|w\rangle|x\rangle|y\rangle|10\rangle=|w\rangle|x\rangle|y\rangle|10\rangle$.
  \item[(4)] If $w=x$ and $w=y$,  $C_U|w\rangle|x\rangle|y\rangle|b\rangle=C_a^{-1}C_bC_a|x\rangle|x\rangle|x\rangle|10\rangle\\=C_a^{-1}C_b|x\rangle|x\rangle|x\rangle|11\rangle=e^{i\theta'}C_a^{-1}|x\rangle|x\rangle|x\rangle|11\rangle=e^{i\theta'}|x\rangle|x\rangle|x\rangle|10\rangle$.
\end{description}
Thus, the proposition has been proved.~$\Box$

In Propostion~4, we have proved that the circuit $C_U$ can implement near-trivial transformation on computational basis states. Finally, we will prove this result for general $n$-qubit states. Without loss of generality, we restrict our attention to pure state.

{\bf Theorem~4:} $C_U(|\varphi\rangle\otimes|x,y,b\rangle)=([x,y,\theta,\theta']|\varphi\rangle)\otimes|x,y,b\rangle$, for any $x,y\in\{0,1\}^n$, where $b=10$, $|\varphi\rangle$ is arbitrary $n$-qubit states and $\theta,\theta'\in[0,2\pi)$ are certain fixed angles.

{\bf Proof:} While $x,y\in\{0,1\}^n$ and $x\neq y$, any $n$-qubit state $|\varphi\rangle$ could be expressed as follows:
\begin{equation}|\varphi\rangle=\eta_x|x\rangle+\eta_y|y\rangle+\sum_{\substack{w\in\{0,1\}^n\\w\neq x,w\neq y}}\eta_w|w\rangle,\end{equation}
where $\sum_{w\in\{0,1\}^n}\left\vert\eta_w\right\vert^2=1$.

For any $n$-qubit state $|\varphi\rangle$, we will have the following derivation from Proposition~4.
\begin{align}
&C_U|\varphi\rangle|x\rangle|y\rangle|10\rangle \nonumber\\
=&C_U\left(\eta_x|x\rangle|x\rangle|y\rangle+\eta_y|y\rangle|x\rangle|y\rangle+\sum_{\substack{w\in\{0,1\}^n\\ w\neq x,w\neq y}}\eta_w|w\rangle|x\rangle|y\rangle\right)|10\rangle \nonumber\\
=&\left(\eta_x\left(cos\theta|x\rangle+sin\theta|y\rangle\right)+\eta_y\left(-sin\theta|x\rangle+cos\theta|y\rangle\right)\right)|x\rangle|y\rangle|10\rangle \nonumber\\
&+\left(\sum_{w\neq x,w\neq y}\eta_w|w\rangle\right)|x\rangle|y\rangle|10\rangle\nonumber\\
=&([x,y,\theta]|\varphi\rangle)|x\rangle|y\rangle|10\rangle.
\end{align}

So, this circuit could implement arbitrary near-trivial rotation $[x,y,\theta]$, for any $x,y\in\{0,1\}^n$, where $x\neq y$ (Note that $\theta$ is fixed here).

While $x,y\in\{0,1\}^n$ and $x=y$, any $n$-qubit state $|\varphi\rangle$ could be expressed as follows:
\begin{equation*}|\varphi\rangle=\eta_x|x\rangle+\sum_{w\in\{0,1\}^n,w\neq x}\eta_w|w\rangle,\end{equation*}
where $\sum_{w\in\{0,1\}^n}\left\vert\eta_w\right\vert^2=1$.

For any $n$-qubit state $|\varphi\rangle$, we will have the following derivation from Proposition~4.
\begin{align}
C_U|\varphi\rangle|x\rangle|x\rangle|10\rangle=&C_U\left(\eta_x|x\rangle|x\rangle|x\rangle|10\rangle+\sum_{w\in\{0,1\}^n,w\neq x}\eta_w|w\rangle|x\rangle|x\rangle|10\rangle\right)\nonumber\\
=&\left(\eta_xe^{i\theta'}|x\rangle+\sum_{w\neq x}\eta_w|w\rangle\right)|x\rangle|x\rangle|10\rangle\nonumber\\
=&\left([x,x,\theta']|\varphi\rangle\right)|x\rangle|x\rangle|10\rangle.
\end{align}

Thus, this circuit can implement arbitrary near-trivial rotation $[x,x,\theta']$, for any $x\in\{0,1\}^n$ (Note that $\theta'$ is fixed here).
Since $[x,y,\theta,\theta']=[x,y,\theta]$, if $x\neq y$ and $[x,y,\theta,\theta']=[x,x,\theta']$, if $x=y$, the theorem is proved.~$\Box$

Until now, we have verified that the quantum circuit $C_U$ in Figure~\ref{fig12} implements arbitrary near-trivial transformation $[x,y,\theta,\theta']$ for any $x,y\in\{0,1\}^n$ (It does not matter whether $x$ is equal to $y$ or not. $\theta$ and $\theta'$ are certain fixed angles).

\begin{figure} [htbp]
%\vspace*{13pt}
\centerline{\epsfig{file=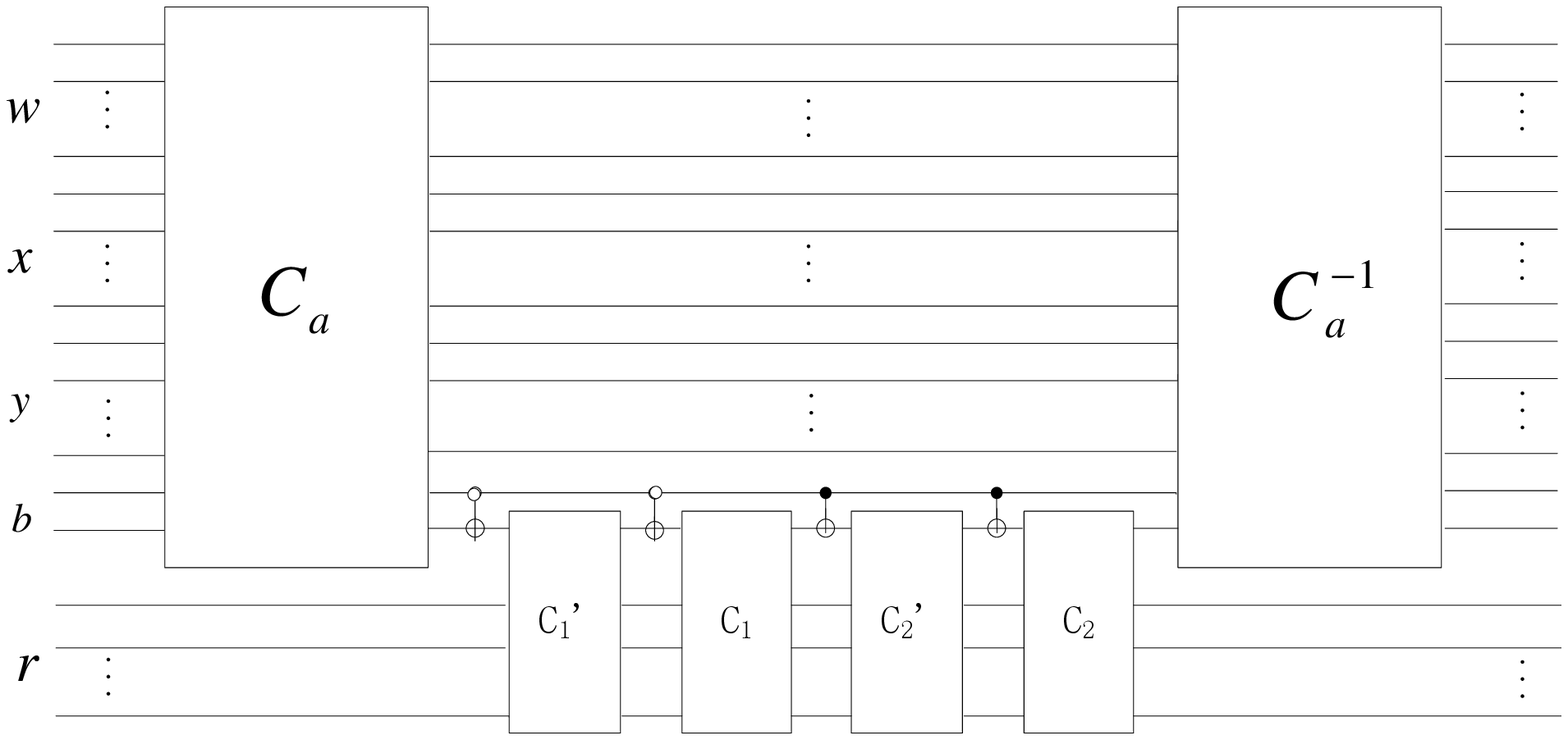, width=10cm}} %100 percent
\vspace*{5pt}
\caption{\label{fig13}Universal quantum circuit $C_U'$ approximately implementing near-trivial transformation.
In Figure~\ref{fig12}, we replace the circuit $C_b$(in the middle part of the circuit)by the quantum circuit $C_b'$ in Figure~\ref{fig11} which is an
approximation of $C_b$, and then we obtain the circuit $C_U'$ which implements near-trivial transformation approximately.
The quantum register $r$ is a $m$-qubit register, and the value of $m$ is determined by the accuracy.}
\end{figure}

According to the analysis in Proposition~3, we could replace the circuit $C_b$ in Figure~\ref{fig12} by the quantum circuit $C_b'$ in Figure~\ref{fig11} which is an approximate implementation. Thus, we obtain the following universal quantum circuit $C_U'$(in Figure~\ref{fig13}) implementing approximately arbitrary near-trivial transformation $[x,y,\theta,\theta']$, for any $x,y\in\{0,1\}^n$ and for any $\theta,\theta'\in(0,2\pi)$. It can be seen that the following two statements hold:

If $x\neq y$,  $C_U'\left(|\varphi\rangle|x\rangle|y\rangle|10\rangle|r\rangle\right)=\left([x,y,0.r\cdot2\pi]|\varphi\rangle\right)|x\rangle|y\rangle|10\rangle|r\rangle$;

If $x=y$, $C_U'\left(|\varphi\rangle|x\rangle|x\rangle|10\rangle|r\rangle\right)=\left([x,x,0.r\cdot4\pi]|\varphi\rangle\right)|x\rangle|x\rangle|10\rangle|r\rangle$.

%%\vspace*{8mm}
%\begin{center}
%\label{fig13}
%\centerline{\psfig{figure=fig13.eps,width=10cm}}\end{center}
%\vspace*{-8mm} {\footnotesize {\bf Figure 13}\quad Universal quantum circuit $C_U'$ approximately implementing near-trivial transformation.
%In Figure 12, we replace the circuit $C_b$(in the middle part of the circuit)by the quantum circuit $C_b'$ in Figure 11 which is an
%approximation of $C_b$, and then we obtain the circuit $C_U'$ which implements near-trivial transformation approximately.
%The quantum register $r$ is a $m$-qubit register, and the value of $m$ is determined by the accuracy.}% \vspace*{8mm}
\clearpage

{\bf Remark~2:}

(a) All the gates in the above circuit (in Figure~\ref{fig13}) are fixed gates without parameters, in other words, the circuit is in standard form of quantum circuit.

(b) This circuit is a polynomial-size quantum circuit.

(c) The exponential number of near-trivial transformations can be implemented in this single quantum circuit.

\section{Discussions}
Quantum circuit is a very important model in the theory and experiment of quantum computing. Quantum circuit model is much simple than QTM model, but is not exactly equivalent to QTM model. For example, given a quantum circuit, its running time is finite, definite and is independent of the inputs. However, the running time of a given QTM dynamically changes and may be infinite while the QTM do not halt on a given input. Indeed, the quantum circuit model has more advantages than QTM model in many aspects. QTM model is the most important tool in studying quantum complexity theory. Our results on quantum circuit present a basis for the further study of QTM.

In Section~4, we have constructed a universal quantum circuit
implementing arbitrary near-trivial transformation. The circuit
$C_U'$ in Figure~\ref{fig13} consists of only some generalized-CNOT and
single-qubit rotations in \\ $\{R_z(\pm\frac{\pi}{2^j}),
R_y(\pm\frac{\pi}{2^j}), j\in{\bf N}\}$. Since stationary normal
form QTM for these quantum gates can be easily constructed, and
according to Bernstein and Vazirani(Lemma 4.9
in~\cite{Bernstein97}), we can dovetail these QTMs one by one and
obtain a stationary normal form QTM which could exactly simulate
this circuit $C_U'$. Thus, we can get a stationary normal form QTM
approximately implementing any near-trivial transformation. It can
be seen that the QTM we described is space-bounded
QTM~\cite{Watrous03}, so we propose a way to construct space-bounded QTM
simulating any quantum circuit. These results will be given in another paper.
In addition, according to the structure of universal quantum Turing machine (UQTM)~\cite{Bernstein97}, near-trivial transformation is an elementary component of UQTM. We believe that near-trivial transformation is fundamental in the model of quantum computation. So it is meaningful to study near-trivial transformation and construct universal quantum circuit for it.

At last, we compare the result of Sousa and Ramos~\cite{Sousa07} with ours in two aspects: 1) the implementation of general unitary transformation, 2) the implementation of certain quantum algorithm described with CNOT and single-qubit gates.
From the first aspect, in order to use the universal cell of Sousa and Ramos~\cite{Sousa07}, the unitary transformation must be decomposed through the following steps: 1) decompose the unitary transformation into several two-level unitary transformation according to the way of Deutsch~\cite{Deutsch89,Ekert96}, 2) decompose the two-level unitary transformation into generalized Toffoli gates and controlled unitary gates in the way of DiVincenzo~\cite{DiVincenzo95}, 3) decompose Toffoli gates into CNOT and single-qubit gates, and decompose the controlled unitary gates according to ABC decomposition~\cite{Nielsen02}. Through the three steps, we obtain a quantum circuit consisting of only CNOT and single-qubit gates. Base on this, several universal cells can be applied to the implementation of the unitary transformation. However, we construct universal cell directly proceeding from two-level unitary transformation which can be seen as the product of a near-trivial rotation and a near-trivial phase shift. In this way, it is not required to decompose the two-level unitary transformation.
However, from the second aspect, the realization of certain quantum algorithm described with CNOT and single-qubit gates means only simulating a quantum circuit consists of CNOT and single-qubit gates. This problem has been well solved by Sousa and Ramos~\cite{Sousa07}. In this case, it is better to use their universal cell.

\section{Conclusions}
\noindent We implement near-trivial transformation with a universal quantum circuit rather than a family of quantum circuits. This quantum circuit is constructed by means of only CNOT and single-qubit rotations, and its size is polynomial. Any near-trivial transformation could be encoded, then the encoding string and the quantum data are both inputs of the universal quantum circuit. The circuit has a result consisting of the encoding string and the result of the near-trivial transformation. It can be seen that, the exponential number of near-trivial transformations can be implemented using a single circuit in our construction. Since any unitary transformation
can be decomposed into a product of some near-trivial transformations, our result may contribute to the design of universal quantum computer.

\section*{Acknowledgements}
This work was supported by the National Natural Science Foundation of China under Grant No.60573051.

%% References with bibTeX database:

%\bibliographystyle{model1c-num-names}
%\bibliography{<your-bib-database>}

%% Authors are advised to submit their bibtex database files. They are
%% requested to list a bibtex style file in the manuscript if they do
%% not want to use model1c-num-names.bst.

%% References without bibTeX database:

\end{document}